\documentclass[a4paper,12pt]{article}

\usepackage{hyperref}

\usepackage{amsmath}
\usepackage{graphicx}

\usepackage[margin=1.0in]{geometry}
\usepackage{authblk}

\newcommand{\dd}{\mathrm{d}}
\newcommand{\ee}{\mathrm{e}}
\newcommand{\ii}{\mathrm{i}}

\title{Note on the charged boson stars\\ with torsion-coupled field}

\author[1]{D.~Horvat}
\author[1]{S.~Iliji\'c\footnote{Email: {\tt sasa.ilijic@fer.hr}}}
\author[2]{A.~Kirin}
\author[1]{Z.~Naran\v ci\'c}

\affil[1]{University of Zagreb,
          Faculty of Electrical Engineering and Computing,
          Department of Applied Physics, Unska 3, HR-10\,000 Zagreb, Croatia}

\affil[2]{Karlovac University of Applied Sciences,
          Ivana Me{\v s}trovi{\'c}a 10, HR-47\,000 Karlovac, Croatia}

\date{\today}

\begin{document}

\maketitle 


\begin{abstract}
Within the framework of the extended teleparallel gravity,
a new class of boson stars has recently been constructed
by introducing the nonminimal coupling
of the scalar field to the torsion scalar.
An interesting feature of these static, spherical,
self-gravitating configurations of the massive complex scalar field
is their central region with outwardly increasing energy density,
surrounded by a thick shell within which the joining
with the usual asymptotically Schwarzschild tail takes place.
In this work we extend the original model with the $U(1)$ gauge field
and we find that the combined effect of the charge
and coupling of the field to torsion
leads to a significant increase of the maximal mass and the particle number
that can be supported against gravity.
We also show that some charged configurations
preserve the property of having the outwardly increasing energy density
over the central region,
regardless of the fact that charging the configurations
affects the anisotropy of the pressures
in the opposite way relative to that of the field-to-torsion coupling terms.
\end{abstract}


\section{Introduction \label{sec:intro}}

Our motivation for considering
the nonminimal coupling of the scalar field to the torsion scalar,
and boson stars in particular,
comes from the recently revived interest
in the theories of gravity based on torsion,
rather than curvature \cite{aldrovandi,maluf}.
It has been known for a long time
that formulating the gravitational action with the torsion scalar $T$,
instead of the curvature scalar $R$ as in the standard theory,
leads to the theory of gravity that is equivalent to general relativity (GR)
and that is known as the teleparallel equivalent of GR (TEGR).
This theory uses the curvature-free Weitzenb\"ock connection
and its dynamical variable is the tetrad---%
a set of four orthonormal fields
that select the local Lorentz frame at every spacetime point.
The extensions of the TEGR are particularly appealing because they are
in some aspects simpler than the similar extensions of GR.
For example, the $f(T)$ theory,
which could be compared with the $f(R)$ extension of GR,
leads to equations of motion that are of the second order,
whereas $f(R)$ leads to equations of the fourth order.
This is due to the fact that the torsion scalar
involves only the first order derivatives of the tetrad,
while the curvature scalar involves the second order derivatives
of the spacetime metric.
However, there are also complications
associated with the extensions of the TEGR
that are not present in the similar extensions
of the standard curvature-based theory.
Within the TEGR there is the six-fold infinity of acceptable tetrads,
related among themselves by rotations and by Lorentz boosts,
which reflects the Lorentz invariance of the theory.
In the simplest extensions of the theory,
such as the $f(T)$ theory \cite{barrow1,barrow2},
or theories involving nonminimal coupling of the scalar field
to the torsion scalar $T$
\cite{harkolobo,scalartorsiondynfeat,scalartorsionderivatives},
the Lorentz invariance of the resulting equations of motion is being lost.
The equations of motion must be engaged to single out the right tetrad,
i.e.\ one must deal with the six degrees of freedom of the tetrad
(related to local rotations and Lorentz boosts),
for which there is no universal procedure.
Regardless of these difficulties,
in several special cases that include
the static spherical symmetry and homogeneous isotropic geometry,
the preferred tetrads leading to physically meaningful
equations of motion have been found \cite{tamanini}
and the solutions that are distinct from
their extended GR counterparts could be constructed and studied.

In the static spherical symmetry,
boson stars are a good starting point
to study the effects of newly introduced terms,
such as those due to scalar-to-torsion coupling,
because they are arguably the simplest mathematical models
of static self-gravitating objects
with the underlying matter model derived directly from the field theory.
Within the standard GR,
boson stars were first constructed by Kaup \cite{KaupKleinGordonGeon}
and by Ruffini and Bonazzola \cite{RuffiniBonazzola}
as self-gravitating configurations of the massive complex scalar field.
The basic model was soon extended with the inclusion
of the field self interaction \cite{CoShaWa86},
electrical charge \cite{jetzercharged},
nonminimal coupling \cite{vanderbij}, etc.
This broad subject has been reviewed
in \cite{liddlemadsen,jetzer,schunck,liebling}.
Boson stars with nonminimal coupling of the scalar field
to the curvature scalar revealed exotic properties
such as configurations with negative principal pressures
in the interior of the star \cite{horvatmarunovic}
or the formation of the photon spheres
in the strong gravitational lensing regime \cite{photonspheres}.
Nonminimal coupling of the scalar field to the torsion scalar
revealed another interesting property---the outwardly increasing
energy density over the interior of the star \cite{torsionbosonstar}.
Such behavior has not been encountered in any other stellar model
with energy-momentum derived from a simple field theory.
In this work we intend to study this feature more closely
and also inspect the behavior of the anisotropy of the principal pressures,
which is a mechanism behind many of the
exotic spherically symmetric solutions in general relativity, e.g.\
wormholes \cite{VisserWHbook} or gravastars
(stars with negative pressures in the center) \cite{visseranisograva}.
We also extend the torsion-coupled model
by adding the local $U(1)$ gauge invariance, or the electric charge,
a component that has not yet been considered in this context.
As the static scalar field and the electrical field
are both sources of the pressure anisotropy
in the static spherically symmetric setting,
we are interested to investigate
how the outwardly increasing energy density 
obtained by scalar-to-torsion coupling will
respond to the inclusion of the electrical charge.

The paper is organized as follows.
A necessary minimum of the technical background on the teleparallel gravity,
mainly restricted to establishing the notation,
is given in Sec.~\ref{sec:bckg}.
Static spherical symmetry is introduced in Sec.~\ref{sec:sss},
and in Sec.~\ref{sec:num} we construct a number of specific
boson star configurations and examine the behavior
of the energy-momentum components.
We discuss our results in Sec.~\ref{sec:kraj}.
Some additional technical details are given in the appendix.
Geometrized units ($c=1=G_{\mathrm{N}}$) are used in the paper.


\section{Charged scalar field coupled to torsion\label{sec:bckg}}

The tetrad fields are denoted by $h_a{}^\mu$,
where the latin index runs over the Lorentz frame coordinates,
and the greek index runs over the usual spacetime coordinates.
The tetrad satisfies the orthonormality relations
$\eta_{ab} = h_a{}^{\mu} h_b{}^{\nu} g_{\mu\nu}$,
$g_{\mu\nu} = h^a{}_{\mu} h^b{}_{\nu} \eta_{ab}$,
$h^a{}_{\mu} h_b{}^{\mu} = \delta^a_b$
and $h^a{}_{\alpha} h_a{}^{\beta} = \delta^{\beta}_{\alpha}$,
where $\eta_{ab}=\mathrm{diag}(-,+,+,+)$ is the metric in the Lorentz frame,
and $g_{\mu\nu}$ is the spacetime metric tensor.
Torsion-based teleparallel gravity
uses the curvatureless Weitzenb\"ock connection
that is defined in terms of $h_a{}^\mu$ as
   \begin{equation} \label{eq:weitz}
   \tilde \Gamma^{\alpha}{}_{\beta\gamma}
      \equiv h_a{}^{\alpha} \partial_{\gamma} h^a{}_{\beta}.
   \end{equation}
The torsion tensor is defined as 
   \begin{equation} \label{eq:ttensor}
   \tilde T^{\alpha}{}_{\beta\gamma} \equiv
   \tilde \Gamma^{\alpha}{}_{\gamma\beta}
        - \tilde \Gamma^{\alpha}{}_{\beta\gamma},
   \end{equation}
and the torsion scalar as
   \begin{equation} \label{eq:tscalar}
   \tilde T \equiv
   \frac14 \tilde T_{\alpha\beta\gamma} \tilde T^{\alpha\beta\gamma}
     + \frac12 \tilde T_{\alpha\beta\gamma} \tilde T^{\gamma\beta\alpha}
     - \tilde T_{\alpha\gamma}{}^{\alpha} \tilde T^{\beta\gamma}{}_{\beta}.
   \end{equation}
The essential property of $\tilde T$
is its relation to the curvature scalar $R$,
   \begin{equation} \label{eq:tr}
   R = - \tilde T - \frac{2}{h} \partial_{\mu} (h \tilde T^{\nu\mu}{}_{\nu}) ,
   \end{equation}
where $h=\det[h^a{}_\nu]=\sqrt{-\det[g_{\alpha\beta}]}$,
showing that the two scalars differ only by a total divergence term.
This implies that while the action of the TEGR,
   \begin{equation} \label{eq:tegraction}
   S_{\mathrm{TEGR}} = \int \mathrm{d} x^4 \, h
       \left( - \frac{\tilde T}{2k} + {\mathcal{L}}_{\mathrm{matter}} \right) ,
   \end{equation}
is not invariant with respect to Lorentz transformations of the tetrad
(i.e.\ the choice of the tetrad),
the resulting equations of motion are equivalent to those of GR
because the variation of the action is not affected by the divergence term.

Here, we consider the action $S = \int \mathcal L \, h \, \dd^4 x$
with the Lagrangian density
   \begin{equation} \label{eq:lagdens}
   \mathcal L =
         \left( \frac1{2k} + \xi \phi^* \phi \right) ( - \tilde T )
       - \frac14 F_{\mu\nu} F^{\mu\nu}
                  - \frac12 g^{\mu\nu} \big( (D_\mu \phi)^* D_\nu\phi
                                            + D_\mu\phi (D_\nu \phi)^* \big)
                  - \mu^2 \phi^*\phi,
   \end{equation}
which involves the massive complex scalar field $\phi$
nonminimally coupled to $\tilde T$ and the gauge field $A_\mu$.
The quantities $k=8\pi$, $\xi$ and $e$ are the coupling constants
and $\mu$ is the field mass.
$F_{\mu\nu} = \partial_\mu A_\nu - \partial_\nu A_\mu$
is the usual field-strength tensor
and $D_{\mu} = \partial_\mu - \ii e A_\mu$ is the gauge-covariant derivative.
Variation of the action with respect to the tetrad
leads to the equation of motion
that can be written in the form of the Einstein equation,
$G_{\mu\nu} = k T_{\mu\nu}$,
where $G_{\mu\nu} = R_{\mu\nu} - \frac12 R g_{\mu\nu}$ is the Einstein tensor
and $T_{\mu\nu}$ is the energy-momentum tensor.
The energy-momentum tensor can be written as
   \begin{equation} \label{eq:emtensor}
   T_{\mu\nu} = \frac{ T^{(0)}_{\mu\nu}
              - 2 \xi S_{\mu\nu}{}^{\alpha} \partial_{\alpha} (\phi^*\phi)
           }{ 1 + 2k \xi \phi^*\phi }
   \end{equation}
where
   \begin{alignat}{1}
   T^{(0)}_{\mu\nu} & = F_{\mu\alpha} F_{\nu}{}^{\alpha}
                    + (D_\mu\phi)^* D_\nu \phi
                    + D_\mu\phi (D_\nu \phi)^* \notag \\
                    & \qquad - g_{\mu\nu} \left(
                          \frac14 F_{\alpha\beta} F^{\alpha\beta}
                        + \frac12 g^{\alpha\beta} \big(
                              (D_\alpha\phi)^* D_\beta \phi
                            + D_\alpha\phi (D_\beta \phi)^* \big)
                        + m^2 \phi^*\phi \label{eq:emtensor0}
                      \right)
   \end{alignat}
is the familiar energy-momentum tensor
that one obtains in the case of the minimal coupling of the scalar field,
either to $R$ in the standard theory,
or to $\tilde T$ within the torsion-based TEGR.
The denominator of (\ref{eq:emtensor})
is the same as in the case of nonminimal coupling
of the scalar field to the curvature scalar,
while the second term in the numerator reveals the difference.
The tensor $S_{\alpha\beta\gamma}$ that appears in this term
is known as the modified torsion tensor
and it can be written in terms of the torsion tensor (\ref{eq:ttensor}) as
   \begin{equation} \label{eq:modtorsion}
   \tilde S_{\alpha\beta\gamma} \equiv
   \frac12 (
       \tilde T_{\beta\alpha\gamma}
     + \tilde T_{\gamma\beta\alpha}
     + \tilde T_{\alpha\beta\gamma}
   ) + g_{\alpha\beta} \, \tilde T_{\gamma}
     - g_{\alpha\gamma} \, \tilde T_{\beta} .
   \end{equation}
The variation of the action with respect to the scalar field gives
   \begin{equation} \label{eq:scalareom}
   \frac1h \partial_\mu (h D^\mu \phi)
   = \ii e A_\mu D^{\mu} \phi
      + (\xi \tilde T + m^2 + 2\lambda \phi^*\phi ) \phi,
   \end{equation}
and the variation with respect to the gauge field gives
   \begin{equation} \label{eq:gaugeeom}
   \frac1h \partial_\mu (h F^{\mu\nu})
   = \ii e \big( (D^\nu \phi) \phi^* - (D^\nu \phi)^* \phi \big).
   \end{equation}
The Lagrangin density (\ref{eq:lagdens}) also implies the conserved current
$j_\mu = \ii \big( (D_\mu \phi)^* \phi - D_\mu \phi \, \phi^* \big)$
with the particle number $N$ and the charge $Q=eN$ as the conserved quantities.

\section{Spherical symmetry and anisotropy of pressures \label{sec:sss}}

In order to study the static spherically symmetric configurations
of the charged scalar field we adopt the usual coordinates
$(t,r,\vartheta,\varphi)$, where $r$ is the area radius,
and we write the line element as
   \begin{equation} \label{eq:ds2}
   \dd s^2 = - \ee^{2\Phi(r)} f^2(r)\, \dd t^2
             + f^{-2}(r) \, \dd r^2
             + r^2 \, \dd\Omega^2  ,
   \end{equation}
where $\dd\Omega^2$ is the line element on the unit 2-sphere,
and $\Phi$ and $f$ are the $r$-dependent metric profile functions.
For the scalar field we take the time-stationary Ansatz
   \begin{equation} \label{eq:phiansatz}
   \phi = \phi(r) \, \ee^{-\ii \omega t},
   \end{equation}
and from here on $\phi$ will denote only the real $r$-dependent part,
and $\phi'$ will denote its radial derivative.
For the gauge field we assume
   \begin{equation}
   A_{\mu} = \delta_\mu^t \psi(r),
   \end{equation}
i.e.\ the field has only the radially dependent electrostatic potential $\psi$.
In the static spherical symmetry
the energy-momentum tensor has the form
$T_\alpha^\beta=\mathrm{diag}(-\rho,p,q,q)$,
where $\rho$ is the energy density
and $p$ and $q$ are the radial and the transverse pressures.
In the case of minimal coupling ($\xi=0$)
these quantities follow from (\ref{eq:emtensor0}) as
   \begin{alignat}{1}
   \rho^{(0)} & = \mu^2 \phi^2 + f^2 \phi'^2
                + f^{-2} \ee^{-2\Phi} (\omega+e\psi)^2
                + \frac12 \ee^{-2\Phi} \psi'^2, \label{eq:rho0} \\
   p^{(0)}    & = - \mu^2 \phi^2 + f^2 \phi'^2
                + f^{-2} \ee^{-2\Phi} (\omega+e\psi)^2
                - \frac12 \ee^{-2\Phi} \psi'^2, \\
   q^{(0)}    & = - \mu^2 \phi^2 - f^2 \phi'^2
                + f^{-2} \ee^{-2\Phi} (\omega+e\psi)^2
                + \frac12 \ee^{-2\Phi} \psi'^2. \label{eq:q0}
   \end{alignat}
In order to include the contribution to the energy-momentum tensor
due to nonminimal coupling to the torsion scalar
one needs the modified torsion tensor.
The tetrad must be chosen first and we adopt the form
   \begin{equation} \label{eq:tetrad}
   h^a{}_{\mu} = \left( \begin{array}{cccc}
   \ee^{\Phi} f & 0 & 0 & 0 \\
   0 & f^{-1} \sin\vartheta \cos\varphi
     &  r \sin\vartheta \cos\varphi & -r \sin\vartheta \sin\varphi \\
   0 & f^{-1} \sin\vartheta \sin\varphi
     &  r \sin\vartheta \cos\varphi &  r \sin\vartheta \cos\varphi \\
   0 & f^{-1} \cos\vartheta & -r \sin\vartheta             &  0
   \end{array}
   \right),
   \end{equation}
which has proven to lead to physically meaningful equations of motion
in the context of static spherical symmetry \cite{boeh1}.
A lengthy but straightforward calculation,
which involves the computation of the Weitzenb\"ock connection (\ref{eq:weitz}),
the torsion tensor (\ref{eq:ttensor}),
and the modified torsion tensor (\ref{eq:modtorsion}),
leads to the following expressions
for the energy density and the pressures:
   \begin{alignat}{1}
   \rho & = \frac{ \rho^{(0)} + 8\xi f (f-1) \phi \phi' / r}{
                   1 + 2 k \xi \phi^2 }, \label{eq:rho} \\
   p & = \frac{ p^{(0)} }{ 1 + 2k\xi \phi^2 }, \\
   q & = \frac{ q^{(0)} - 4 \xi f ( f - 1 + rf' + r f \Phi' ) \phi \phi' / r}{
                                                 1 + 2 k \xi \phi^2 },
   \label{eq:q}
   \end{alignat}
where $\rho^{(0)}$, $p^{(0)}$ and $q^{(0)}$
are given by (\ref{eq:rho0})-(\ref{eq:q0}).
The anisotropy of pressures, which we define as the difference
between the transverse and the radial pressure, follows as
   \begin{equation} \label{eq:aniso}
   q - p =
      \frac{ - 2 f^2 \phi'^2 + \ee^{-2\Phi} \psi'^2
    - 4 \xi f ( f - 1 + r f' + r f \Phi' ) \phi \phi'/r}{ 1 + 2k\xi \phi^2 } .
   \end{equation}
It is easy to see that in the absence of the electric charge ($e=0$, $\psi=0$)
and minimal coupling ($\xi=0$), the anisotropy reduces to
$q-p=- 2 f^2 \phi'^2$ which is negative,
except at $r=0$ and as $r\to\infty$ where it is expected to vanish.

It is also straightforward to express the equations of motion
(\ref{eq:scalareom}) and (\ref{eq:gaugeeom})
in terms of our metric and field Ansatze,
but as the resulting expressions are somewhat lengthy,
we only show the expression used to compute the particle number,
   \begin{equation}
   N = \int j^0 \sqrt{-g} \; \dd^3 x
     = 8 \pi \int_0^{\infty} r^2
       \ee^{-\Phi} f^{-2} \phi^2 (\omega + e \psi) \, \dd r
   \end{equation}
($Q=eN$ is the charge).
The total mass of the configuration is given by the familiar expression
$M = 4\pi \int_0^{\infty} r^2 \rho(r)\,\dd r$,
where $\rho$ is the energy density given by (\ref{eq:rho}).

\section{Properties of the charged torsion-boson stars \label{sec:num}}

As the solutions to the equations of motion
are not accessible in a closed form,
we proceed with constructing them numerically.
Solutions can be obtained by solving a boundary value problem
for the metric profile functions $\Phi$ and $f$,
the fields $\phi$ and $\psi$, and the eigenvalue $\omega$,
with the boundary conditions specified at the center of symmetry
($r=0$) and at the spatial infinity ($r=\infty$).
At $r=0$ we require $\phi=\phi_0$, $\Phi'=0$ and $f=1$,
while at $r=\infty$ we require $\phi=\psi=0$.
These boundary conditions,
together with the Einstein equation which provides
three independent differential equations,
and the equations of motion for the fields
(\ref{eq:scalareom}) and (\ref{eq:gaugeeom}),
fully determine the solution.
The value of $\phi_0$ (scalar field amplitude at $r=0$)
that enters through the boundary condition
is used to parametrize the family of solutions
corresponding to the fixed values of the field mass $\mu$
and the coupling constants $e$ (electric charge)
and $\xi$ (nonminimal coupling of the scalar field to the torsion scalar).
For a chosen set of the values of $\mu$ $e$ and $\xi$
we can compute a spectrum of solutions
starting from a very small value of $\phi_0$.
(We have used $\mu=1$ with no loss of generality,
see \cite{jetzercharged} for the rescaling scheme that can be applied.
We have also limited our analysis to solutions
without nodes in the scalar field.)
As $\phi_0$ increases, the particle number and the mass of the configuration
increase as well, until they simultaneously reach the respective maxima.
The configurations with maximal $N$ and $M$ are known as critical
because it in the case of the minimal coupling of the field
they mark the transition from the dynamically stable
to the dynamically unstable configurations.
This behavior of $N$ and $M$
is shown in Fig.~\ref{fig:one} for four sets of parameters.
The critical configurations are indicated with letters A to D,
and their most important parameters are listed in Table {\ref{tbl:one}.
As it is clear from the data in the table,
we have constructed the configurations corresponding to
the minimally coupled electrically neutral field (model A),
minimally coupled electrically charged field (model B),
nonminimally coupled electrically neutral field (model C),
and the critical configuration
with the nonminimally coupled electrically charged field (model D).
One can see that introducing the charge alone
makes it possible to support a larger number of particles
against gravity and obtain a stable boson star
of larger mass \cite{jetzercharged}.
The same holds for introducing of the field-to-torsion coupling,
while the combined effect of charge and torsion coupling
leads to a still larger increase in the critical $N$ and $M$.

\begin{figure}
\begin{center}
\includegraphics{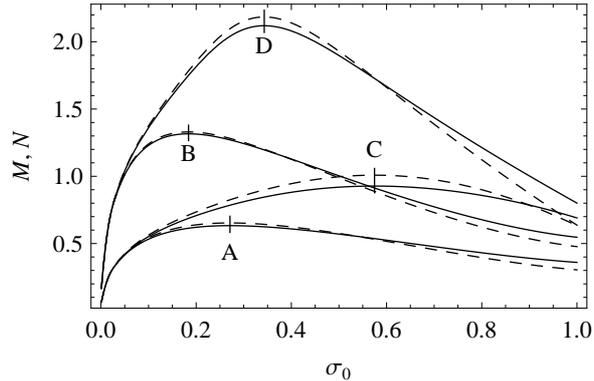} \quad
\caption{Mass (solid lines in units $M_{\mathrm{Pl}}^2/\mu$)
and particle number (dashed lines in units $M_{\mathrm{Pl}}^2/\mu^2$)
of boson stars with parameters given in Table \ref{tbl:one}
and for a range of values of $\sigma_0=k^{-1/2}\phi(0)$.
Critical configurations are indicated with vertical lines and letters.
\label{fig:one}}
\end{center}
\end{figure}

\begin{table}
\begin{center}
\caption{Parameters of the four critical configurations of the boson stars.
Models C and D include nonminimal coupling of the scalar field
to the torsion scalar.\label{tbl:one}}
\begin{tabular}{cccccccc}
Model
& $\mu$
& $e$
& $\xi$
& $\sigma_0$
& $\omega$
& $N/(M_{\mathrm{Pl}}^2/\mu^2)$
& $M/(M_{\mathrm{Pl}}^2/\mu)$ \\
\hline
A & 1 & 0 & 0 & 0.271 & 0.853 & 0.653 & 0.633 \\
B & 1 & 3 & 0 & 0.184 & 0.947 & 1.330 & 1.316 \\
C & 1 & 0 & 8 & 0.575 & 0.685 & 1.008 & 0.927 \\
D & 1 & 3 & 8 & 0.343 & 0.890 & 2.184 & 2.119
\end{tabular}
\end{center}
\end{table}


The model A represents the critical configuration
of the minimally coupled electrically neutral field.
%
%
Since the earliest works on the boson stars,
this has been the common reference configuration
when considering extensions of the simple model.
The energy density and the pressures are shown
in the upper left panel of Fig.~\ref{fig:two}.
As is expected on the basis of (\ref{eq:aniso}),
the anisotropy of the pressures is negative everywhere
except at $r=0$ where it must vanish by virtue of symmetry
and at spatial infinity where both $p$ and $q$ vanish asymptotically.
The transverse pressure crosses zero at finite $r$
and asymptotically approaches zero from the negative regime as $r\to\infty$.


The model~B includes the electric charge.
The value of the coupling constant $e=3$
is chosen close to the maximal value that the configuration can support.
The upper right panel of Fig.~\ref{fig:two} shows that,
on the scale set by the central energy density of the configuration,
the anisotropy of pressures is of small amplitude.
However, this in no way means that its role is unimportant.
The anisotropy of pressures in the interior is negative,
which is the characteristic feature of the minimally coupled field,
while in the asymptotic tail of the configuration it becomes positive.
This reflects the fact that, as $r\to\infty$,
the metric is asymptotically Reissner--Nordstr\"om,
which implies $\rho=-p=q=Q^2/(8\pi r^4)$, $Q$ being the electric charge.
In this model the radial pressure changes sign
and is negative in the asymptotic tail,
which is exactly the opposite of what we had in model~A,
and is a characteristic of electrically charged configurations.
The coupling constant $e$ (charge) can be further increased,
but as one approaches the maximal amount of charge the boson star can support,
the anisotropy becomes positive over the whole configuration.


Model~C includes the nonminimal coupling of the scalar field
to the torsion scalar, but no electric charge.
The value of the field-to-torsion coupling constant $\xi=8$
has been chosen large enough for the configuration to develop
the outwardly increasing energy density over the interior.
The behavior of pressure anisotropy is exactly the opposite
of what we had in model~B.
In the interior we have the positive anisotropy
and in the asymptotic tail it becomes negative.
The transverse pressure has a zero crossing,
while the radial pressure remains positive.
If one used a larger value of $\xi$
the above properties would only be further amplified,
but would not change qualitatively.


In model~D we include both the electric charge
and the nonminimal coupling of the scalar field to the torsion scalar.
The values of the coupling constants are those
that were used individually in models B and C.
The combined effect of charge and torsion
preserves the outwardly increasing energy density over the interior and
results in the positive anisotropy of pressures throughout the boson star.
The radial pressure has the zero crossing.
This is exactly the opposite of what one finds in a simple boson star.
As in model~B, the coupling constant $e$ could be further increased,
but as it approaches the maximal (critical) value of the charge
that can be supported by the boson star,
the outwardly increasing energy density changes to outwardly decreasing,
which means that the configuration becomes
qualitatively similar to model~B.


\begin{figure}
\begin{center}
\includegraphics{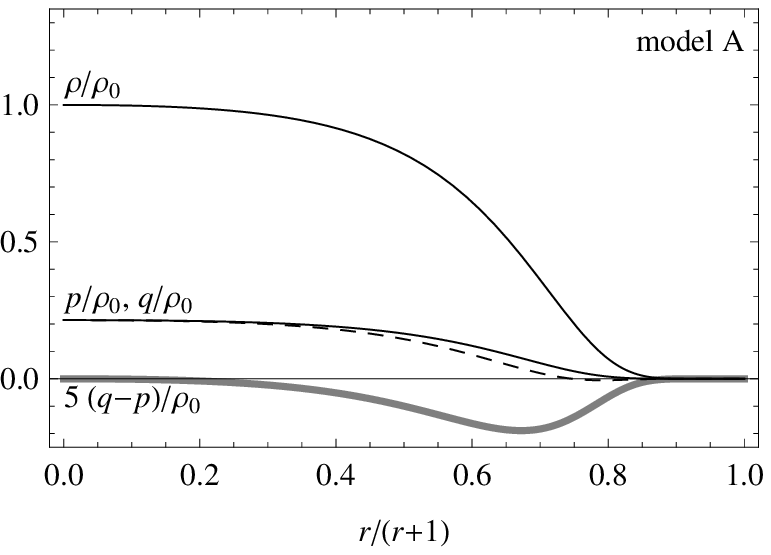}
\includegraphics{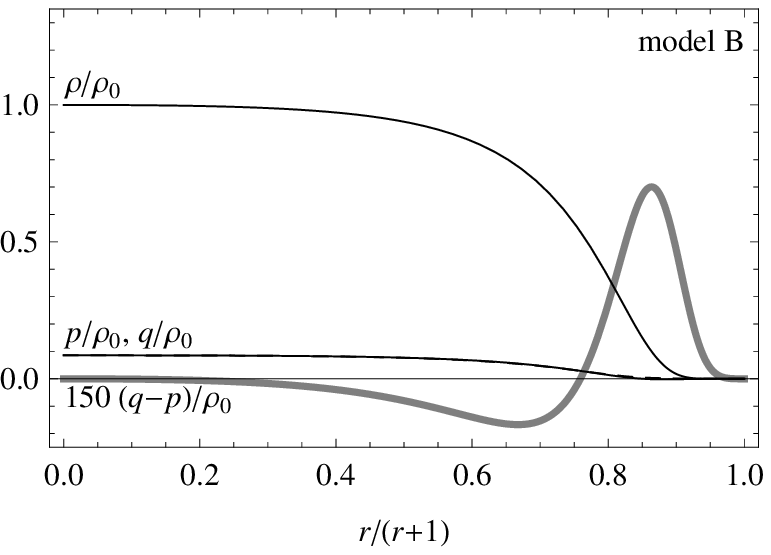} \\
\includegraphics{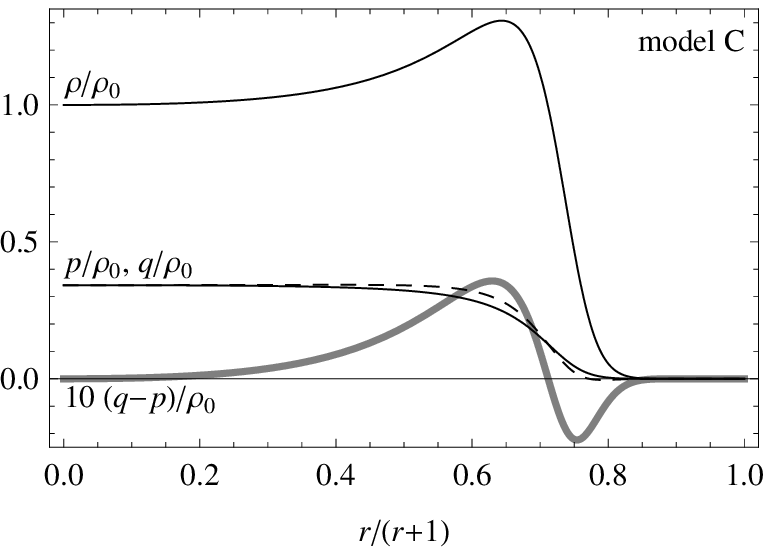}
\includegraphics{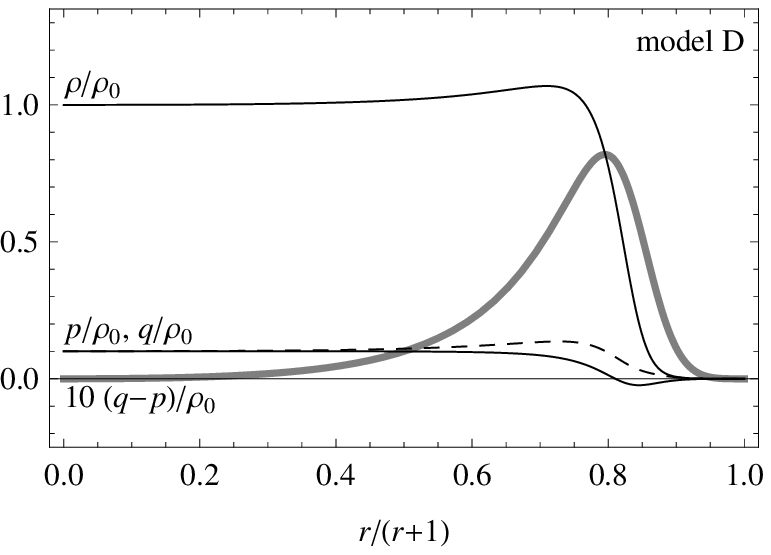}
\caption{Energy-momentum components of the critical configurations A--D
(see Table~\ref{tbl:one} for parameters): energy density (solid thin line)
relative to the central energy density $\rho_0$,
radial and transverse pressure (solid and dashed pair of thin lines)
relative to $\rho_0$, and the anisotropy (gray thick line)
relative to $\rho_0$ and scaled up by a factor for clarity. \label{fig:two}}
\end{center}
\end{figure}

\section{Discussion and conclusions \label{sec:kraj}}

In the static spherically symmetric setting,
the scalar field produces the negative anisotropy of principal pressures
(defined as the transverse minus the radial pressure),
while the radial electrostatic field
gives a positive contribution to the pressure anisotropy.
These two components contribute in the opposite directions,
so one can speak of the scalar
and of the electrostatic sign of the anisotropy.
The matter model that we considered also includes
the coupling of the scalar field to the torsion scalar,
which is the third mechanism contributing to the pressure anisotropy,
and also the component giving raise
to the outwardly increasing energy density in some configurations.
From the expression for the anisotropy (\ref{eq:aniso}) it is not obvious
in what way the torsion-coupling terms affect the anisotropy.
In addition, there is no direct way to isolate the mechanism
responsible of generating the outwardly increasing energy density.
We therefore obtained numerical evidence that shows the trends
that take place when the three sources of anisotropy are active simultaneously.
A small amount of the charge added to a simple boson star
changes the sign of the anisotropy from the negative (scalar)
to positive (electromagnetic) regime only in the outer layer,
i.e.\ in the asymptotic tail,
while as the amount of charge approaches the critical value
beyond which the charged star does not exist,
the anisotropy becomes positive throughout the configuration.
The nonminimal coupling of the field to torsion, but without charge,
changes the sign of the anisotropy into positive (electromagnetic)
only in the interior part of the configuration,
while the anisotropy remains negative (scalar) in the asymptotic tail.
If both charge and torsion coupling are present simultaneously,
provided that field-to-torsion coupling and charge are not too weak,
we find that the anisotropy is positive throughout the configuration
and also the outwardly increasing energy density over the
interior of the boson star.
The energy density returns to the usual outwardly decreasing behavior
only as the charge approaches the maximal amount that can be supported.
The unusual property of outwardly increasing energy density
is here shown to be present in a wider range of physical situations
than it was known previously \cite{torsionbosonstar}.
Therefore the torsion-coupled scalar field confirms itself
as an interesting candidate for the modeling self-gravitating structures.
It is also worth noting that torsion coupling of the scalar field
helped support a larger number of particles against gravity,
which is a trend similar to that of the charge \cite{jetzercharged}
or of the quartic field self interaction \cite{CoShaWa86}.

Before closing, we should not forget to mention
that the matter model we considered
was constructed within the framework of the extended teleparallel gravity,
which means that,
in order to work with physically meaningful equations of motion,
a tetrad suitable for the assumed geometry had to be chosen first.
As we considered the static spherical symmetry,
we could use the good tetrad provided by \cite{tamanini}.
However, applications of this matter model
outside of the static spherical symmetry or homogeneous isotropic geometry
could be hampered by the difficulties in finding the suitable tetrad.
For example, the tetrad compatible with the
time-dependent spherically symmetric geometry seems to be still out of reach.

\vskip 1em \noindent {\bf Acknowledgments:}
This work is supported by the University of Zagreb
Grant No.~VIF2014-PP2-11.
Partial support comes from ``NewCompStar,'' COST Grant No.~MP1304.


\appendix

\section{Explicit equations of motion and the numerical technique}

For completeness here we give the explicit form of some expressions that are,
in order to keep the presentation compact,
not included in the main body of the paper.

Adopting the metric (\ref{eq:ds2}), the Ricci or curvature scalar reads
\begin{equation}
R = R^\mu{}_\mu = \frac{2}{r^2} - 2f'^2
  - \frac{2f}r \big( f' (4+3r\Phi') + r f'' \big)
  - \frac{2f^2}{r^2} \big( (1+r\Phi')^2 + r^2 \Phi'' \big),
\end{equation}
and the three nontrivial components of the Einstein tensor
$G^\mu{}_\nu = R^\mu{}_\nu - \frac12 R \delta^\mu_\nu$ are
\begin{alignat}{1}
G^t{}_t & = \frac1{r^2} \big( - 1 + f^2 + 2 r f f' \big) \label{eq:eintentt},\\
G^r{}_r & = \frac1{r^2} \big( -1 + f^2(1+2r\Phi') + 2 r f f' \big),\\
G^\vartheta{}_\vartheta = G^\varphi{}_\varphi & =
f'^2 + \frac{f}{r}\big( f' (2+3r\Phi') + r f'' \big)
+ \frac{f^2}{r} \big( \Phi' + r \Phi'^2 + r \Phi'' \big).
\label{eq:eintenphiphi}
\end{alignat}
The Einstein equations, $G^\mu{}_\nu = k T^\mu{}_\nu$,
are obtained by combining (\ref{eq:eintentt})--(\ref{eq:eintenphiphi}),
and the components of the energy-momentum tensor,
$T^{\mu}{}_\nu = \mathrm{diag}(-\rho,p,q,q)$,
that are given by
(\ref{eq:rho})--(\ref{eq:q}) and (\ref{eq:rho0})--(\ref{eq:q0}).
Adopting the tetrad (\ref{eq:tetrad}),
the torsion scalar (\ref{eq:tscalar}),
which is related to the curvature scalar by (\ref{eq:tr}), reads
\begin{equation}
T = - \frac2{r^2} (1 - f)(1 - f - 2r ( f' + f \Phi' ) ) .
\end{equation}
The equation of motion for the scalar field (\ref{eq:scalareom})
can be written as
\begin{alignat}{1}
\phi'' = & - \frac{\ee^{-2\Phi} \phi (\omega + e \psi)^2 }{ f^4 }
           - \frac{\phi'(2rf'+f(2+r\Phi'))}{rf}
           + \frac{\phi \mu^2}{f^2} \notag \\
 & \mbox{} - \frac{2\xi\phi(1-f) ( 1 - f - 2r(f'+f\Phi') )}{r^2 f^2} ,
\end{alignat}
where the terms that are specific to the scalar-to-torsion coupling
are grouped in the lower line.
The only nontrivial component of the equation of motion
for the gauge field (\ref{eq:gaugeeom}) is its time component.
It does not involve terms due to scalar-to-torsion coupling and it reads
\begin{equation}
\psi'' = \frac{2 e \phi^2 (\omega + e \psi)}{f^2}
       - \frac{\psi'(2 - r \Phi')}{r}.
\end{equation}
Our procedure for solving
the system of nonlinear ordinary differential equations
for the four unknown functions,
$f$, $\Phi$, $\phi$ and $\psi$, 
subject to the appropriate boundary conditions (see Sec. \ref{sec:num}),
consists of changing the radial variable $r\to x=r/(r+1)$
in order to formulate the problem on the compact domain
and of solving the resulting boundary value problem (BVP) numerically.
The frequency $\omega$
that appears in the harmonic Ansatz (\ref{eq:phiansatz}) for the scalar field
has the role of the eigenvalue and is obtained for every converged solution.
We used the collocation algorithm
implemented in the numerical code COLSYS \cite{colsys}.
In order to double-check the validity of the numerical solutions,
all of them were confirmed by the independent initial value integration
that can be easily carried out once the eigenvalue $\omega$
has been determined by solving the BVP.


\providecommand{\href}[2]{#2}\begingroup\raggedright\endgroup

\end{document}